\documentclass[11pt,letter]{article}

\usepackage{chicago}
\usepackage{amsmath}
\usepackage{graphicx}

\begin{document}
\noindent
REPLICATION AT PERIODICALLY CHANGING MULTIPLICITY OF INFECTION PROMOTES
  STABLE COEXISTENCE OF COMPETING VIRAL POPULATIONS
\bigskip

\noindent
Claus O. Wilke$^1$, Daniel D. Reissig$^2$, and Isabel S. Novella$^2$
\bigskip

\noindent
{\em
$^1$Digital Life Laboratory, California Institute of Technology, Mail Code
136-93, Pasadena, California 91125, USA. wilke@caltech.edu\\
$^2$Department of Microbiology and Immunology, Medical College of Ohio,
Toledo, Ohio 43614, USA. isabel@mco.edu
}
\bigskip

\newpage

\begin{abstract}
  RNA viruses are a widely used tool to study evolution experimentally. Many
  standard protocols of virus propagation and competition are done at
  nominally low multiplicity of infection (m.o.i.), but lead during one
  passage to two or more rounds of infection, of which the later ones are at
  high m.o.i. Here, we develop a model of the competition between wild type
  (wt) and a mutant under a regime of alternating m.o.i. We assume that the
  mutant is deleterious when it infects cells on its own, but derives a
  selective advantage when rare and coinfecting with wt, because it can profit
  from superior protein products created by the wt.  We find that, under these
  assumptions, replication at alternating low and high m.o.i.\ may lead to the
  stable coexistence of wt and mutant for a wide range of parameter settings.
  The predictions of our model are consistent with earlier observations of
  frequency-dependent selection in VSV and HIV-1. Our results suggest that
  frequency-dependent selection may be common in typical evolution experiments
  with viruses.
\end{abstract}


\bigskip

RNA virus populations grown on cell culture in the laboratory are often
considered as single-niche systems, and are used to test basic evolutionary
theories, such as Muller's ratchet, the evolution of recombination, or the
potential costs of host radiation
\shortcite{Chao90,Duarteetal92,Escarmisetal96,Chaoetal97,Yusteetal99,TurnerElena2000}.
In a single-niche system, mutation pressure is the only source of
polymorphisms in the population, while polymorphisms can also be maintained by
negative frequency dependent selection if several niches are available.
Frequency-dependent selection in RNA viruses grown \emph{in vitro}
\shortcite{Elenaetal97,TurnerChao99,Yusteetal2002,TurnerChao2003} demonstrates
that multiple niches can be available even in these simple laboratory systems.

Frequency-dependent selection among viruses grown \emph{in vitro} is typically
caused by complementation, that is, within-cell interactions between different
virus strains. When several viruses coinfect the same cell, they share genetic
material and protein products while they replicate. This type of interaction
can for example lead to the accumulation of defective interfering particles
(DIPs) \shortcite{BanghamKirkwood90,Szathmary92,Frank2000}, virus particles
that cannot replicate by themselves because they lack essential genes. DIPs
can coexist with non-defective virus particles because they complement their
defective genomes with genes from the non-defective particles when both
coinfect the same cell. Other effects caused by within-cell interactions are
phenotypic mixing and hiding
\shortcite{NovickSzilard51,Brenner57,Huangetal74,Hollandetal89,WilkeNovella2003}
or recombination and reassortment
\shortcite{Kingetal82,Lai92,Whiteetal95,Rodriguezetal98,SteinhauerSkehel2002}.

If complementation is the main cause for frequency-dependent selection, then
fitness should be frequency-dependent only at high multiplicity of infection,
when every cell is infected by many virions at once. At low multiplicity of
infection, virions rarely have to share a cell, and within-cell interactions
are not expected to play a dominant role. Indeed, ~\shortciteN{TurnerChao2003}
reported frequency-dependent fitness at multiplicity of infection (m.o.i.) of
5, but did not observe frequency-dependent fitness at m.o.i.\ of 0.002, in
strains of bacteriophage $\Phi$6 that have been selected for cooperation or
defection during coinfection.  Seemingly at odds with these considerations are
results reported by \shortciteN{Elenaetal97} on vesicular stomatitis virus
(VSV).  Elena et al.\ found frequency dependent selection and stable
coexistence of Fpop40 and wild type in experiments carried out at an m.o.i.\
of 0.1.  The probability with which a cell is infected with more than one
particle at this m.o.i.\ is less than 0.005, so that complementation seems to
be too rare to have a major impact on the selection dynamics. However,
within-cell interactions are only rare at the beginning of an infection cycle:
Most viruses, including VSV, are conventionally cultured by inoculation of a
cell monolayer at low m.o.i., followed by incubation until maximum titers are
reached \shortcite{Hollandetal91}. At an m.o.i.\ of 0.1, only 10\% of the host
cells get infected upon inoculation.  These cells start to release virus
progeny (approximately 4 hours post infection for VSV), and this virus progeny
infects the remaining uninfected cells. Since a single cell can produce
thousands of infectious particles, the cells that were initially uninfected
will get infected by a considerable number of virus particles.  Therefore, the
m.o.i.\ during that last part of the infection is high, and complementation
should be common.  Results similar to those of Elena et al.\ were also found
by \shortciteN{Yusteetal2002} in experiments with HIV, and the same
explanation applies.

In this communication, we present a mathematical model and experimental
results which incorporate the two independent rounds of infection that VSV
undergoes during a typical 24h passage, one at low m.o.i.\ and one at high
m.o.i. We analyze the model and demonstrate that stable coexistence of two
mutants is indeed supported by complementation during coinfection, and thus
provide a biological mechanism that explains the results of
\shortciteN{Elenaetal97} and \shortciteN{Yusteetal2002}.

\bigskip

\centerline{
\sc Materials and Methods
}

\centerline{
\it Cells and Viruses
}

Vesicular stomatitis virus (VSV) is a member of the family
\emph{Rhabdoviridae}, and infects insects and mammals
\shortcite{RoseWhitt2001}. It is a negative-sense single-stranded RNA virus,
and its genome has a length of approximately 11,000 nucleotides, coding for at
least five genes.  We have employed two populations of Indiana serotype
(Mudd-Summers strain). Wild type (wt) is our reference laboratory strain, to
which we have assigned a fitness of 1.0.  MARM N is an monoclonal antibody
resistant mutant of low fitness.  MARM N was generated by selection of a clone
from wt that was resistant to the I1 Mab, followed by eight small
plaque-to-small plaque passages and two rounds of amplification in BHK-21
cells at low m.o.i.  The host cells were BHK-21 obtained from John Holland's
laboratory, and I1-hybridoma cells were a kind gift of Douglas Lyles
\cite{LefrancoisLyles82}.  Cell growth methods have been described in detail
elsewhere \shortcite{Hollandetal91}.
\bigskip

\centerline{
\it Fitness Assays
}

Fitness was determined by competition of wt and MARM populations
\shortcite{Hollandetal91}. Regular fitness assays were done by mixing wt and
MARM at equal ratios and using the mixture to infect a monolayer at a m.o.i.\
of 0.1 in T-25 flasks (with a surface of 25 cm$^2$).  Viral progeny after
20-24 hours of replication was appropriately diluted, and used to infect a
fresh monolayer at an m.o.i.\ of 0.1. This process was repeated for up to 3
competition passages.  The original mixture and the viral yield produced after
each competition passage were titrated by triplicate plaque assay in the
presence and absence of I1-Mab to calculate the ratio wt:MARM N.  Changes in
the log of normalized ratios were plotted against passage number to obtain the
slope of the linear fit, which is the fitness value.  To carry out
competitions that encompassed a single round of infection we used T175 flasks
(175 cm$^2$ of surface).  Low m.o.i.\ competitions (0.1 PFU/cell) were done as
indicated above, except that viral yield was harvested at 9 hours, to avoid
sampling of second generation progeny.  We also carried out competitions at
high m.o.i.\ (i.e.\ 10 PFU/cell).  For consistency the viral progeny was also
harvested at 9 hours post-infection.  Because passages at high m.o.i.\ promote
the accumulation of defective interfering particles
\shortcite{DePoloHolland86,DePoloetal87}, competitions were not taken past the
first competition passage.  Each fitness value was determined between 9 and 13
times.
\bigskip

\centerline{
\it Estimation of Combined Fitness
}

We estimated the combined fitness of MARM N undergoing one round of infection
at m.o.i.=0.1 and one round of infection at m.o.i.=10 by multiplying the
respective fitness values obtained from the 9h passages. That is, if $w_{\rm
  l}$ is the relative fitness of the mutant at low m.o.i., and $w_{\rm h}$ is
the relative fitness of the mutant at high m.o.i., then the combined fitness
of the mutant $w_{\rm c}$ is $w_{\rm c}=w_{\rm l}\times w_{\rm h}$. We
obtained the error on $w_{\rm c}$ from Gaussian error propagation: If
$\sigma_{\rm l}$ is the error for $w_{\rm l}$, and $\sigma_{\rm h}$ is the
error for $w_{\rm h}$, then the error $\sigma_{\rm c}$ of the combined fitness
follows as $\sigma_{\rm c}^2 = w_{\rm l}^2\sigma_{\rm l}^2+w_{\rm h}^2\sigma_{\rm h}^2$.
\bigskip

\centerline{
\sc Model
}

We consider the competition between wild type (wt) and a debilitated mutant
strain. We assume that at low m.o.i., wt replicates faster than the mutant, so
that the mutant has a fitness of $1-s_1$ in comparison to wt. At high m.o.i.,
we assume that wt and mutant interact, and that the fitness of mutant relative
to wt is dependent on the relative frequency $x$ of wt in the population.  We
write the mutant's fitness relative to wt as $1-c(x)$.

After a single round of infection at low m.o.i., the frequency $x$ of
wt becomes  $f_{\rm low}(x) = x/[x+(1-s_1)(1-x)]$.
Likewise, after a single round of infection at high m.o.i., the
frequency $x$ of wt becomes $f_{\rm high}(x) = x/\{x+[1-c(x)](1-x)\}$.
We assume that a typical 24h passage of VSV consists of one round of infection
at low m.o.i. and one round of infection at high m.o.i. Therefore, the
frequency of wt after one passage becomes (we denote by $x'$ the frequency of
wt at the end of the passage) $x' =  f_{\rm high}[f_{\rm low}(x)]$.

For the function $c(x)$, we assume a linear dependence on $x$, $c(x) = s_1 -
2s_2x$. With this choice, the mutant strain has the same fitness at low
m.o.i.\ and at high m.o.i.\ when it is on its own, and has, at high m.o.i., an
increasingly higher fitness as the fraction of wt increases in the population.
The idea behind this assumption is that the mutant is defective or less
efficient in some functions, and is able to exploit the wt in these functions
under coinfection.  In particular, if $s_2>s_1/2$, then the mutant, when rare
and coinfecting with wt, manages to produce more progeny than the wt. In this
case, the wild type acts as cooperator and the mutant as defector in the
terminology of \shortciteN{TurnerChao99}.

With the definitions given in the previous two paragraphs, we find
\begin{equation}\label{spec-next-passage}
x' = \frac{[1+s_1(x-1)]x}{1+(x-1)[K+xL]},
\end{equation}
where $K=s_1(3-3s_1+s_1^2)$ and $L=s_1(2s_1-s1^2+2s_2)-2s_2$.  By iterating
Eq.~\eqref{spec-next-passage}, we obtain the change of the wt frequency over
time.
\bigskip

\centerline{
\sc Results
}

\centerline{
\it Standard VSV Passage Consists of Two Rounds of Replication
}

We carried out competition assays between MARM N and wt, allowing only for a
single round of cell infection at different m.o.i.s. We found a significant
difference between the fitness values from 9h assays at different m.o.i.\
(Table 1), while population size did not have an effect
\shortcite{Novellaetal2003}.  We also found a significant difference between
the fitness value from the 24h assays and from the 9h assays at low and high
m.o.i.\ (Table 1). Initial frequency of MARM N had at most a very weak effect
on fitness \shortcite{Novellaetal2003}.  We compared the fitness obtained from
the standard 24h competition assays with the combined fitness of the 9h
assays. We calculated the combined fitness by multiplying the fitness value
for a single round of infection at m.o.i.=0.1 with the fitness value for a
single round of infection at m.o.i.=10. We found that the fitness value from
the 24h competition agreed very well with the combined fitness value from the
two 9h competitions (Table~\ref{MARM-N-fitness}). Therefore, we concluded that
it is reasonable to model standard 24h MARM N passages as a combination of two
rounds of replication, one at low m.o.i.\ and one at high m.o.i.  \bigskip

\centerline{
\it Replication at Alternating M.O.I.\ Can Lead to Stable Coexistence
  of Strains
}

\begin{figure}[h]
\centerline{
\includegraphics[width=.8\columnwidth]{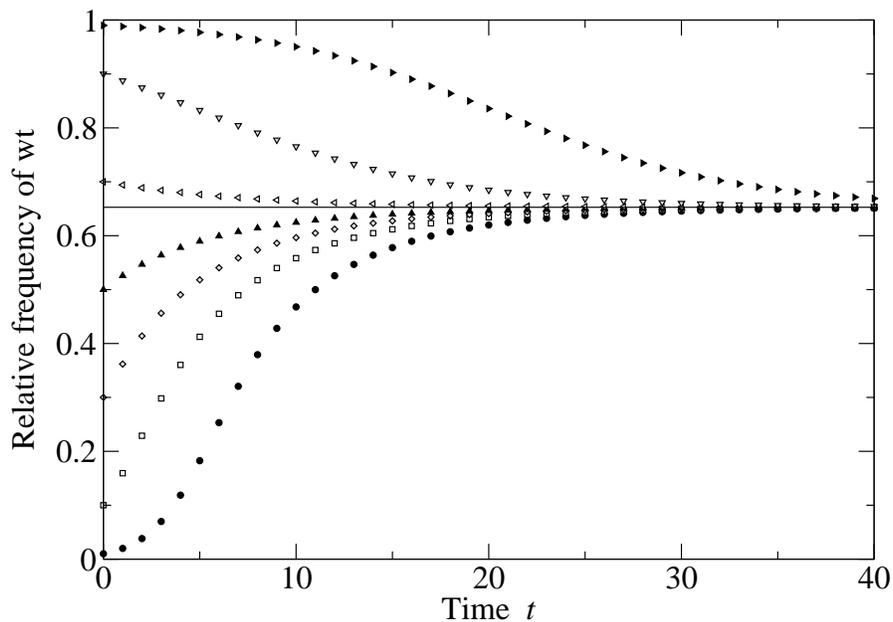}
}
\caption{\label{fig:equil-approach}Approach of a stable polymorphism from
  various initial wt frequencies. Data points were generated by
  iterating Eq.~\eqref{spec-next-passage}. The solid line indicates the
  equilibrium frequency as predicted by Eq.~\eqref{wt-level-equil}. Parameters were $s_1=0.3$, $s_2=0.5$.}
\end{figure}

Numerical simulations of Eq.~\eqref{spec-next-passage} showed that mutant and
wt can stably coexist in our model, and that the equilibrium frequency is
independent of the initial frequency of the wt (see
Fig.~\ref{fig:equil-approach}). We obtained a full characterization of the
possible dynamics of our model from fixed-point analysis. Details are given
in the Appendix. The main conclusions from the fixed-point analysis are the
following: When $s_2$ is sufficiently large, that is, when the mutant derives
a large advantage from coinfecting with the wt, then mutant and wt can stably
coexist. The equilibrium concentration of wt in this case is given by
Eq.~\eqref{wt-level-equil}. If $s_2$ is too small, then wt will always drive
the mutant to extinction. Whether $s_2$ is large or small depends on how it
compares to $s_1$, see Eq.~\eqref{eq:phase-boundary}. For small $s_1$, $s_2$
is large when it is larger than $s_1$.

\begin{figure}[h]
\centerline{
\includegraphics[width=.8\columnwidth]{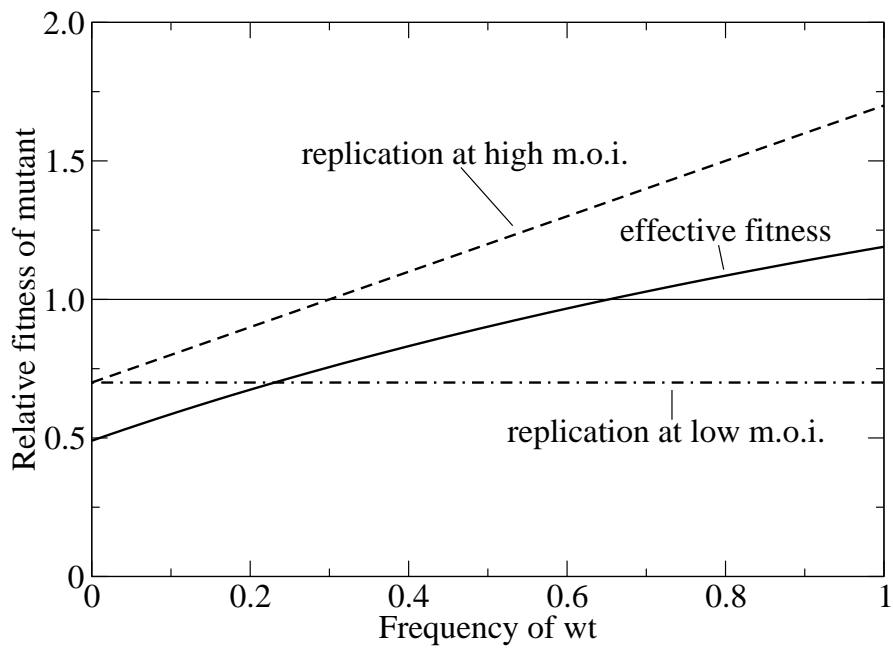}
}
\caption{\label{fig:fitnesses}Relative fitness of mutant as a function of the wt
  concentration during the first and second rounds of replication, and
  effective fitness during a complete passage as determined
  by Eq.~\eqref{effective-fitness}. Parameters were $s_1=0.3$,
  $s_2=0.5$.}
\end{figure}

With linear stability analysis, we could demonstrate that a stable coexistence
between mutant and wt is possible. However, this method is fairly abstract,
and does not help us to understand \emph{why} mutant and wt  coexist. We
can take a more graphical approach to this question by considering the
effective fitness of the mutant in one passage. The total growth of the wt in
one passage is $x'/x$, and that of the mutant is $(1-x')/(1-x)$. Therefore,
the growth of the mutant relative to the wt [and thus the effective fitness
$w(x)$] is
\begin{equation}\label{effective-fitness}
  w(x) = \frac{1-x'}{x'}\frac{x}{1-x} = \frac{1+s_1 x - (K+xL)}{1+s_1(x-1)}\,.
\end{equation}
In general, mutant and wt can stably coexist if the relative fitness of the
mutant grows with the abundance of the wt, and crosses the value 1 for some
positive wt concentration (see for example the discussion by \shortciteNP{TurnerChao2003}). For the case of our model, mutant and wt can therefore
coexist if the effective fitness of the mutant $w(x)$ crosses the value 1 at
some value of $x$. Figure~\ref{fig:fitnesses} illustrates the mutant fitness
during the first and second round of infection, and the effective fitness of
the mutant for the complete passage. We observe that the effective fitness
increases monotonically with the wt concentration, and crosses the value 1 at
the wt concentration $x=0.65$. Therefore, for the parameter values of
Fig.~\ref{fig:fitnesses} ($s_1=0.3$, $s_2=0.5$), the wt grows faster than the
mutant if it is less abundant than 0.65, while the mutant grows faster if the
wt is more abundant than 0.65. Over time, wt and mutant therefore settle into
an equilibrium with wt concentration $x=0.65$. Note that this wt
concentration corresponds to the equilibrium value predicted by
Eq.~\eqref{wt-level-equil}.

From the above considerations, we find that another way to determine the
parameter region in which coexistence is possible is to calculate under which
circumstances $w(1)<1$. This calculation leads to the same condition
Eq.~\eqref{eq:phase-boundary} as the linear stability analysis.
\bigskip

\centerline{
\sc Discussion
}

We have shown that overall fitness, as measured under standard laboratory
conditions, can be described as a composite of two steps of selection.  The
first step is a low m.o.i.\ infection cycle, where selection can freely
operate.  The second step occurs during replication at high m.o.i., when
coinfection promotes complementation, and selection can no longer operate
efficiently on genomes of low fitness. Furthermore, we have shown that the
separation of virus passages into two independent rounds of replication, one
at low m.o.i.\ and one at high m.o.i., can lead to frequency-dependent
selection and stable coexistence of mutants in VSV.  Our results are in very
good qualitative agreement with results in the literature for VSV
\shortcite{Elenaetal97} and also human immunodeficiency virus type 1 (HIV-1)
\shortcite{Yusteetal2002}. The experimental settings in both reports include
infections done at an initially low m.o.i.~(0.1 PFU/cell), and in both cases a
second, high m.o.i.\ infection round takes place.  Both
\shortciteN{Elenaetal97} and \shortciteN{Yusteetal2002} had proposed
within-cell interactions as a potential mechanism contributing to
frequency-dependent selection. Our results suggest that complementation during
the second round of infection at high m.o.i.\ is sufficient to explain
frequency-dependent selection and coexistence observed both in VSV and HIV-1.

Interestingly, the evolutionary regimes followed to generate Fpop40 (i.e.
repeated transmission with large populations at low m.o.i.) would be
equivalent to that followed in phage $\Phi$6 to select cooperators
\shortcite{TurnerChao98,TurnerChao99}.  However, the results show that Fpop40
behaves as an overall defector.  The HIV strains reported by Yuste et al.\
have a history of repeated genetic bottleneck \shortcite{Yusteetal99}.  While
this history would favor the accumulation of deleterious mutations, there is
no reason to assume that such mutations would be implicated in defection.
Defectors in $\Phi$6 were obtained by repeated passages at large population
size and high m.o.i \shortcite{TurnerChao98}.  Thus, both in VSV and HIV-1,
strains that were not a priori selected for defection (and were evolved under
very different protocols) behaved as overall defectors, which implies that
frequency-dependent selection may be a common phenomenon in virus experimental
evolution, and may often occur in unexpected situations.

While frequency-dependent selection could be a fairly common observation in
VSV under standard 24h passages (and in other viruses under analogous
conditions), stable coexistence of two strains is not necessarily common as
well.  Stable coexistence of two strains is only possible if the strain with
lower fitness can exploit the strain with higher fitness at high m.o.i. and
high concentration of the latter strain. However, it seems plausible that many
deleterious strains, even when they profit from coinfection with an
advantageous strain, will at most be able to fare as good as the advantageous
strain. For example, if protein products are freely shared between two strains
in a cell, then both strains will profit equally from the molecular machinery
present in the cell, and will produce equal fractions of offspring virions.
In this situation, the deleterious strain has a disadvantage in the round of
infection at low m.o.i., and is selectively neutral in the round of infection
at high m.o.i., so that its effective fitness remains always below 1.
Coexistence between the two strains is not possible in this case.

Our model can be applied to viruses replicating cytolytically or persistently.
In both cases, natural infections often start with few virions infecting
individual cells, and proceed with multiple infections by virus progeny.
While gene expression may vary between persistent and cytolytic replication,
sharing of protein products in the infected cells can occur in both scenarios.
From the model or the data we cannot infer which step of the viral replication
cycle is responsible for frequency-dependent fitness differences, but there are
steps that can be ruled out.  Entry is not likely to be involved.  Both in VSV
and HIV-1 complementation would not operate during the high-m.o.i. round of
infection \cite{WilkeNovella2003}.  Transcription and translation are also
unlikely to be the cause of coexistence, at least for VSV.  Changes in
transcriptional and/or translational levels would be reflected as changes in
the corresponding protein levels, and be the same for both competitors during
coinfection.  Thus, a deleterious mutant could never reach a fitness value
higher than that of wild type.  Cis-acting replication and encapsidation
signals are the most reasonable candidates to carry mutations involved in
defection, as discussed by \shortciteN{TurnerChao2003}.

In this report, we do not explicitly consider the quasispecies nature of the
strains, and regard the two strains as if they were genetically homogeneous.
This approach has been used successfully in the quasispecies literature
\shortcite{SchusterSwetina88,Wilkeetal2001b,Wilke2001b}, and can be justified
mathematically when mutations from one strain to the other are rare
\shortcite{SchusterSwetina88}. We must interpret our model in the sense that
the properties that we ascribe to the two strains are not properties of
particular genotypes, but rather average properties of the two quasispecies.

The results we have presented here have consequences for
RNA viruses as a model of experimental evolution: The standard mode of
interpretation and modeling of virus evolution experiments is to assume
non-interacting particles. [For example, see the modeling work done by
\shortciteN{Soleetal99} to explain the experimental
results of \shortciteN{Clarkeetal94}, or the modeling work done by
\shortciteN{Rouzineetal2003} to explain the results by \shortciteN{Novellaetal95a} and \shortciteN{Novellaetal99}.] However,
standard experimental protocols allow for more than one infection round, which
means that frequency-dependent selection must always be considered a
possibility in the interpretation of the results. To give meaningful results,
future experimental protocols should either avoid the second round of
infection at high m.o.i., or test explicitly for the presence of frequency
dependence.

\bigskip

\centerline{
\sc Acknowledgments
}

C.O.W. was supported by the NSF under contract No DEB-9981397 to Chris Adami.
I.S.N. was supported by NIH grant AI45686.


Corresponding Editor: Santiago F.\ Elena

\begin{appendix}

\centerline{
\sc Appendix: Linear Stability Analysis
}

We find the fixed points of our model by setting $x'=x$ in
Eq.~\eqref{spec-next-passage} and solving for $x$. There are three fixed
points, at $x=0$ (population consists only of mutant), $x=1$ (population
consists only of wt), and
\begin{equation}\label{wt-level-equil}
x=(s_1-K)/L
\end{equation}
(mixed equilibrium, coexistence between mutant and wt), where
$K=s_1(3-3s_1+s_1^2)$ and $L=s_1(2s_1-s1^2+2s_2)-2s_2$.
Since $x$ is the relative frequency of wt, the fixed point defined by
Eq.~\eqref{wt-level-equil} is meaningful only when $(s_1-K)/L$ falls between 0
and 1.

We begin the linear stability analysis with the fixed point $x=0$. After
inserting $x=\epsilon$ into Eq.~\eqref{spec-next-passage} and expanding to
first order in $\epsilon$, we obtain $x' = \frac{1}{(1-s_1)^2}\epsilon +
O(\epsilon^2)$.  Since $1/(1-s_1)^2>1$, this fixed point is always unstable.
For the fixed point $x=1$, we obtain, after inserting $x=1-\epsilon$ into
Eq.~\eqref{spec-next-passage}, $x' =
1-(1-s_1)(1-s_1+2s_2)\epsilon+O(\epsilon^2)$. This fixed point is stable when
$(1-s_1)(1-s_1+2s_2)<1$, which is equivalent to $s_2<s_1(1-s_1/2)/(1-s_1)$.
Finally, for the fixed point $x=(s_1-K)/L$, we find
$x' = \frac{s_1-K}{L} +
  \Big[(1-s_1)^2+\frac{(2-s_1)^2s_1^2}{2(1-s_1)s_2}\Big]\epsilon
  +O(\epsilon^2)$. This fixed point is stable if
$\Big|(1-s_1)^2+\frac{(2-s_1)^2s_1^2}{2(1-s_1)s_2}\Big| < 1$,
which implies that $s_2>s_1(1-s_1/2)/(1-s_1)$\,. It is straightforward to
verify that the same condition guarantees that the fixed point falls between 0
and 1, and thus is meaningful as a concentration of wt.

To summarize, we find that $x=0$ is unstable, $x=1$ is stable when $(s_1-K)/L$
does not fall between 0 and 1, and unstable otherwise, and $x=(s_1-K)/L$ is
stable when it falls between 0 and 1.  The condition under which $x=(s_1-K)/L$
falls between 0 and 1 is
\begin{equation}\label{eq:phase-boundary}
  s_2>s_1\frac{1-s_1/2}{1-s_1}\approx s_1\,.
\end{equation}
Therefore, wt and mutant can stably coexist if (for small $s_1$) $s_2$ is
larger than $s_1$. In other words, if at high m.o.i. the advantage that
mutant gets from the presence of wt at equal concentrations is at least as
large as the disadvantage of the mutant at low m.o.i., then wt and mutant can
coexist.

\end{appendix}

\cleardoublepage

\begin{table}[h]
 \centerline{
 \begin{tabular}{c|c|c}
  & m.o.i.=0.1 & m.o.i.=10 \\[.2cm]\hline&&\\[-.2cm]
  9h & 0.54$\pm$0.03 & 0.79$\pm$0.06 \\
  24h & 0.43$\pm$0.02 & N/A \\
  2$\times$9h & 0.42$\pm$0.05 & N/A\\
 \end{tabular}
 }
 \bigskip
 \noindent

\caption{Relative fitness of MARM N for 9h and 24h passages at different m.o.i.
The combined fitness value (indicated by 2$\times$9h) was obtained by
multiplying the fitness
values for m.o.i.=0.1 and m.o.i.=10. The error was
calculated using Gaussian error propagation, see Methods section.}\label{MARM-N-fitness}
\end{table}

\end{document}